\documentclass{iaus}
\usepackage{graphics}

\def\hi{{\sc H\,i}}

\def\oiii{{\sc [O\,iii]}}
\def\halpha{H\,$\alpha$}

\title[PNe as Mass Tracers in Galaxies] 
{Planetary nebulae as mass tracers in galaxies}

\author[A.J. Romanowsky]   
{Aaron J. Romanowsky%
}

\affiliation{Departamento de F\'{i}sica, Universidad de Concepci\'{o}n, Casilla 160-C, Concepci\'{o}n, Chile \break email: romanow@astro-udec.cl\\[\affilskip]
}

\pubyear{2006}
\volume{234}
\jname{Planetary Nebulae in our Galaxy and Beyond}
\editors{M. J. Barlow \& R. H. M\'endez, eds.}
\pagerange{1--8}
\date{?? and in revised form ??}
\begin{document}

\maketitle

\begin{abstract}
Planetary nebula are useful kinematic tracers of the stars in all galaxy types.
I review recent observationally-driven developments in the study of galaxy mass profiles.
These have yielded surprising results on spiral galaxy disk masses
and elliptical galaxy halo masses.
A key remaining question is the
coupling between PNe and the underlying stellar populations.

\end{abstract}

\firstsection 
\section{Introduction}

Planetary nebulae (PNe) are of great importance in galaxy studies
as {\it unique} observable proxies for stars in regions of low surface brightness.
They can be used to infer the spatial, kinematic, and chemical properties of the
underlying stellar population, at distances of up to 100 Mpc
(cf. $\sim$1~Mpc for resolved stellar spectroscopy).
Here I review the use of PN kinematics to probe mass distributions in galaxies,
with particular attention to disk masses of spiral galaxies (\S \ref{sec:disk}),
and halo masses of elliptical galaxies (\S \ref{sec:ell}).
PN kinematics are also useful for studying angular momentum, orbit properties,
and substructure in galaxies---topics
beyond the scope of this review.

In all such dynamical studies, it is important to know {\it what is} the underlying
stellar population traced by the observed PNe in external galaxies.
These (bright) PNe are frequently assumed to originate with an
``old'' stellar population that
comprises the bulk of the stellar mass in the galaxy.
This assumption has a great deal of theoretical uncertainty
({\it e.g.}, \cite[Marigo \etal\ 2004]{Marigo04}; \cite[Ciardullo \etal\ 2005]{Ciardullo05};
\cite[Buzzoni \etal\ 2006]{Buzzoni06}), but
empirically, it works remarkably well when PN numbers are compared to
stellar photometry.
I will return to this issue later, especially in \S \ref{sec:iss}.
A summary is in \S \ref{sec:summ}.


\section{Planetary nebula kinematics in disk galaxies}\label{sec:disk}

\subsection{The Milky Way.}
There are unique pros and cons to studying PNe in the Milky Way vs.
extragalactically.
Probably the biggest dynamical challenge is the PN distance uncertainty.
PN rotation curves have been derived to 14~kpc
({\it e.g.}, \cite[Schneider \& Terzian 1983]{Schneider83};
\cite[Maciel \& Lago 2005]{Maciel05}),
supporting the presence of a dark halo.
But it may be more useful to turn the analysis around and use other 
mass constraints to derive the PN distance scale
(\cite[Phillips 2001]{Phillips01});
or to use PNe to study the dynamical structure of the Galaxy
({\it e.g.}, \cite[Beaulieu et al. 2000]{Beaulieu00}).

\subsection{Observations in spiral galaxies}

PN kinematics in spiral galaxies outside the Milky Way have
not been studied much (relative to elliptical galaxies)
because of possible contamination from other emission-line regions,
and because of the availability of alternative mass tracers.
However, there are some recent developments in this area, based
upon observations summarized in Table~\ref{tab:tab1}.

\begin{table}\def~{\hphantom{0}}
  \begin{center}
  \caption{Selected observations of planetary nebula velocities in external galaxies}
  \label{tab:tab1}
  \begin{tabular}{lccccrc}\hline
      Galaxy & Type & $M_B$ & Distance & Telescope+Instrument & Number of& Ref. \\
	& & & (Mpc) & & velocities & \\
\hline
{\it Spirals} &&&&&& \\
M94 & Sab & -19.7 & 5 & WHT+ISIS & 67 & D+00 \\ 
M33 & Sc & -19.0 & 1 & WIYN+HYDRA & 140 & C+04\\
M31 & Sb & -21.2 & 1 &WHT+WYFFOS, WHT+PN.S& 2615 & H+06, M+06 \\
M83 & SBc & -20.5 & 4 & CTIO+HYDRA & 182 & HC06 \\
\hline
{\it Lenticulars}&&&&&&\\
NGC 3384 & E5/S0 & -19.5 & 11 & CTIO+RFP & 50 & TW95, SW06 \\
NGC 5866 & S0 & -20.1 & 14 & WHT+ISIS & 34 & G00 \\
NGC 7457 & E4/S0 & -18.6 & 12 & WHT+PN.S & 100 & in prep \\
\hline
{\it Ellipticals}& {\it (early}&{\it work)}&&&&\\
M32 & E & -16.4 & 1 & Lick+ITS, KPNO+IIDS & 15 & NF86 \\
NGC 3379 & E1 & -19.9 & 10 & KPNO+NESSIE & 29 & C+93 \\
NGC 1399 & E1 & -21.0 & 19 & NTT+EMMI & 37 & A+94 \\
NGC 5128 & E2p & -20.7 & 4 & AAT+RGO, CTIO+ARGUS & 433 & H+95 \\
NGC 4406 & E4 & -21.3 & 16 & NTT+EMMI & 19 & A+96 \\
NGC 1316 & E3p & -21.8 & 20 & NTT+EMMI & 43 & A+98 \\
\hline
{\it Ellipticals}& {\it (recent}&{\it work)}&&&&\\
NGC 4697 & E3 & -19.9 & 11 & UT1+FORS1 & 535 & M+01 \\
NGC 821 & E4 & -20.4 & 22 & WHT+PN.S & 140 & R+03, in prep \\
NGC 3379 & E1 & -19.9 & 10 & WHT+PN.S, CTIO+RFP & 187 & R+03, SW06, in prep \\
NGC 4494 & E1 & -20.4 & 16 & WHT+PN.S & 248 & R+03, in prep \\
NGC 5128 & E2p & -20.7 & 4 & AAT+2dF, CTIO+Argus,Hydra & 780 & P+04 \\
NGC 1344 & E5 & -20.2 & 19 & UT3+FORS1 & 195 & T+05 \\
M87 & E3 & -21.4 & 15 & UT2+FORS2 & 200 & in prep \\
NGC 4472 & E2 & -21.7 & 15 & WHT+WYFFOS, UT2+FORS2 & 80 & in prep \\
\hline
  \end{tabular}
 \end{center}
\end{table}

The standard observational technique for extragalactic PN kinematics
involves an imaging survey with ``off-band'' and ``on-band'' images,
where ``on-band'' usually means a narrow-band filter 
centered around the 5007\AA{} \oiii{} line.
Velocities are acquired by follow-up with a multi-object spectrograph.
Some examples of this are wide-field imaging surveys of M33 and M83,
followed by fiber spectroscopy with velocities accurate to $\sim 5$~km~s$^{-1}$
(\cite[Ciardullo \etal\ 2004]{Ciardullo04};
\cite[Herrmann \& Ciardullo 2006]{Hermann06}).

Another technique is to use a type of slitless spectroscopy called
{\it counter-dispersed imaging} (CDI),
in which PN detection and velocity measurement are done all in one step.
Used occasionally with existing instrumentation
({\it e.g.}, \cite[Douglas \etal\ 2000]{Douglas00}),
CDI has really come into its own with the custom-built
Planetary Nebula Spectrograph 
(PN.S; \cite[Douglas \etal\ 2002]{Douglas02}).
Originally optimized for observation of 5007\AA{} only,
the PN.S has a new arm for simultaneous \halpha{} imaging,
due for commissioning in late 2006.
This addition allows for improved PN detection reliability,
better background contaminant rejection, and
the capability of measuring $F$(\oiii{})/$F$(\halpha{}) line ratios.

A spectacular success of the PN.S has been the wide-field kinematic survey
of the Local Group spiral M31, with 2615 PN velocities
over 7~deg$^2$, encompassing much of the bulge, disk, and halo 
(\cite[Merrett \etal\ 2003, 2006]{Merrett03,Merrett06}).
There is also an extensive traditional imaging+fiber survey
(\cite[Halliday \etal\ 2006]{Halliday06}).
Comparing techniques, the PN.S is clearly more efficient, but
has slightly worse velocity accuracy,
and probably poorer astrometry for facilitating further spectroscopic study.

\subsection{Rotation curves of spiral galaxies}

The most fundamental kinematic property of a spiral galaxy is
the rotation curve, or circular velocity profile, which provides
the cumulative mass profile via the relation
$v_{\rm c}^2(r) =  G M(r)/r$.
It is infeasible to obtain $v_{\rm c}(r)$ 
from integrated-light stellar spectroscopy in the nearest spirals
because of their large angular extent and rapidly decreasing surface brightness
profiles.
With PN kinematics in M31 and M33, $v_{\rm c}(r)$ is obtained to 
4--6 disk scale-lengths $R_{\rm d}$, 
and found to agree with the flat rotation curves obtained from
\hi{} and CO gas measurements
(modulo the asymmetric drift corrections;
\cite[Ciardullo \etal\ 2004]{Ciardullo04};
\cite[Merrett \etal\ 2006]{Merrett06}).
This appears to rule out magnetic fields as an alternative explanation for the flat rotation curves
({\it e.g.}, \cite[Battaner \& Florido 2005]{Battaner05}).

\subsection{Disk masses of spiral galaxies}

Although \hi{} measurements in large samples of spiral galaxies
make it clear that $v_{\rm c}(r)$ is fairly constant at large $r$,
implying large discrepancies between the visible and total masses
(\cite[Persic \etal\ 1996]{Persic96}),
there is considerable freedom for decomposing $M(r)$ into its subcomponents.
Thus there is a strong degeneracy between the mass-to-light ratio $\Upsilon_*$ used for the stars,
and the inferred properties of the dark matter (DM) halo
({\it e.g.}, \cite[Bottema 1997]{Bottema97}).
The degeneracy can be broken by measuring the velocity dispersion of the
stellar disk, especially in the face-on case.
This is evident with the isothermal disk approximation, where the
vertical velocity dispersion $\sigma_z$ is directly proportional to
the disk surface mass density $\Sigma$,
and the scale-height $z_0$ should be fairly constant:
$\sigma_z^2(R)=\upi G \Sigma(R) z_0$.

\begin{figure}
\centering
\resizebox{6.5cm}{!}{\includegraphics{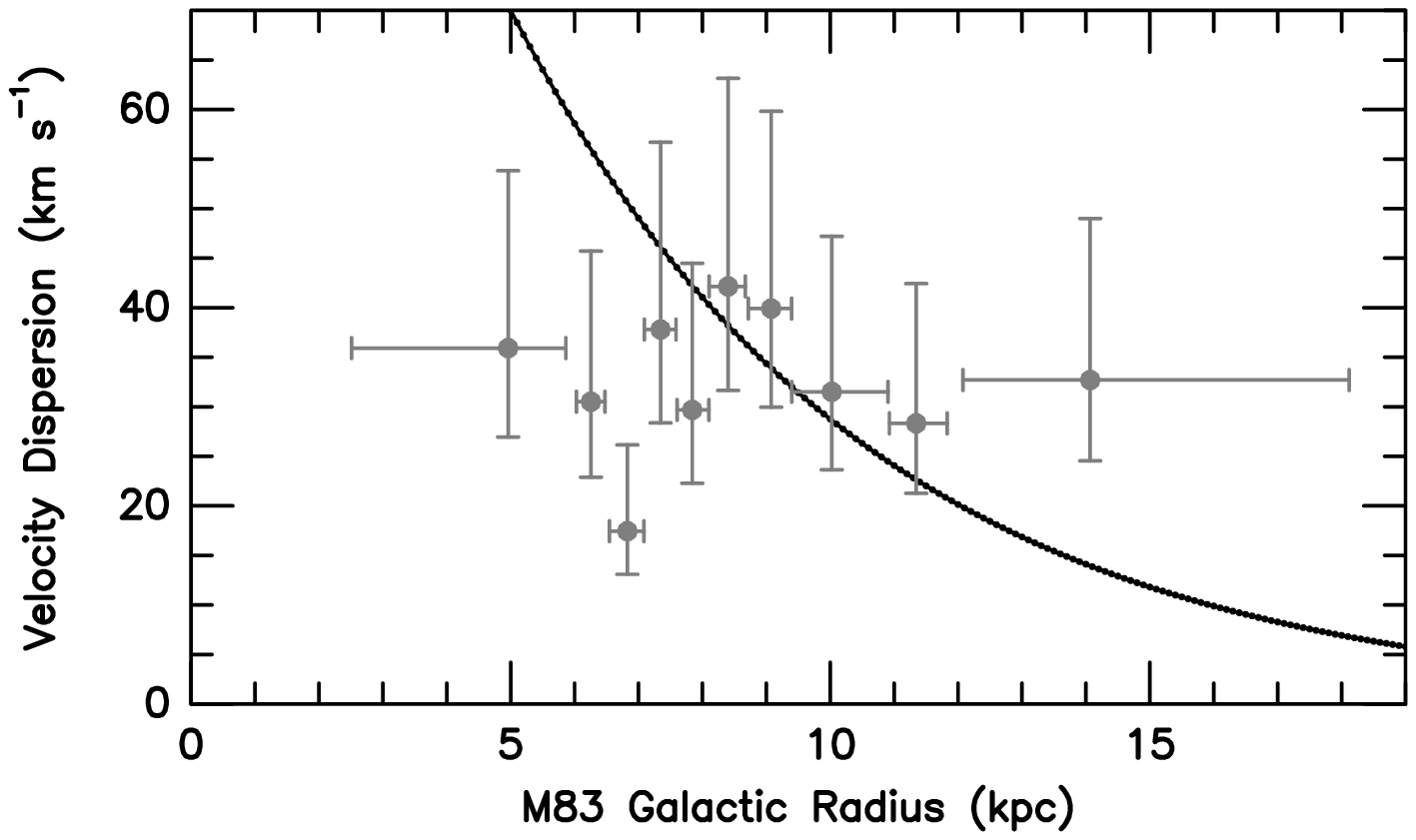} }
\resizebox{6.5cm}{!}{\includegraphics{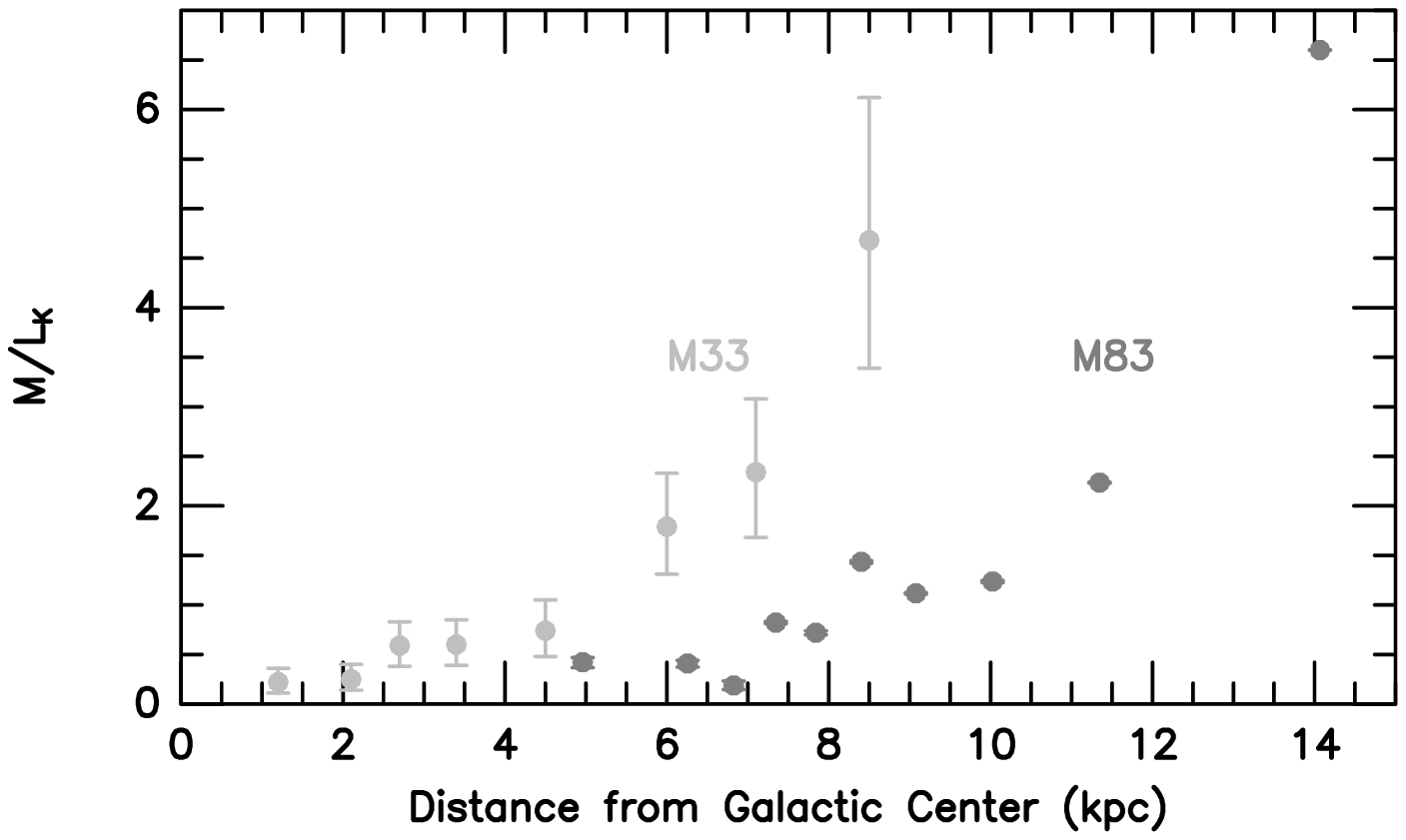} }
\caption[]{
{\it Left:}
Vertical velocity dispersion profile of PNe in M83, as a function of radius.
Error-bars show the data, and curve shows expectation from constant
mass-to-light ratio disk.
{\it Right:}
Implied local $K$-band mass-to-light ratio of disk, with radius, for M33 and M83.
}\label{fig:fig1}
\end{figure}

Such dispersions have been measured in inner disks, but one cannot assume that
$\Upsilon_*$ is the same in the outer disk---and in these low surface brightness regions,
PN kinematics are generally the only accessible stellar tracer.
In M33 and M83, the PN dispersion is found to decrease fairly slowly with radius,
implying a disk mass scale-length much larger than the photometric $R_{\rm d}$
(see Fig.~\ref{fig:fig1}, left; \cite[Ciardullo \etal\ 2004]{Ciardullo04};
\cite[Herrmann \& Ciardullo 2006]{Hermann06}).
This means that $\Upsilon_*$ increases markedly with radius
(Fig.~\ref{fig:fig1}, right)---the opposite
to what one expects from inside-out disk formation.
A similar effect is seen in M31, where warp models are also tried out, but
still don't fit the observations
(\cite[Merrett \etal\ 2006]{Merrett06}).

The M33 PN measurements imply that the rotation curve is dominated by the disk
in the inner parts, apparently ruling out the cuspy dark halo profiles
predicted by $\Lambda$CDM theory.
However, the disk is not massive enough to account for {\it all} the ``missing mass'',
ruling out the cold gas DM theory
(\cite[Pfenniger \etal\ 1994]{Pfenniger04}).
Another alternative is the MOND gravitational theory,
but with this, the M33 PN data point to a self-consistency problem
(\cite[Stubbs \& Garg 2005]{Stubbs05}).
It seems the only major model for spiral galaxy mass profiles which is not ``disproved''
is a DM halo with a shallow core---but further work is clearly needed.

\subsection{Lenticular galaxies}

Lenticular galaxies (S0s) are still mysterious objects.
It is not known whether they are more akin to spiral galaxies
(resulting from gas stripping and quenching), or
to ellipticals (resulting from major galaxy mergers).
These two basic structural-formational scenarios can be
distinguished via their large-radius kinematics---an
observational goal we are now beginning to address with the PN.S
(E. Noordermeer \etal\ in prep).

\section{Planetary nebula kinematics in elliptical galaxies}\label{sec:ell}

Elliptical galaxies are notoriously difficult to model, with their
complicated internal structures and dearth of suitable kinematic tracers.
With PN data, 
there is the potential for engaging in global ``three-dimensional'' dynamical
studies of ellipticals.
For example, on the 4-m William Herschel Telescope,
the IFUs SAURON and OASIS (with AO)
can probe a galaxy's central parts from parsec to kpc scales,
while the PN.S reaches out to tens of kpc.
Such contiguous constraints over four decades in radius are important because
of the coupling between triaxial orbits in the halo and the central supermassive black hole---though
there is an implicit assumption that the PNe and stars share the same dynamics
(see \S\ref{sec:iss}).

\subsection{Observations in elliptical galaxies}

Early work on PNe in elliptical galaxies produced small but tantalizing kinematic samples 
in their halos (see Table~\ref{tab:tab1}).
The exception was the nearby NGC~5128, which remains the best-studied
early-type galaxy.
Recent improvements in instrumentation and methods have led
to an explosion of PN velocities (again, see Table~\ref{tab:tab1}).
Particularly important has been the exploration for the first time
of ordinary $L^*$ ($M_B \sim -20.3$) ellipticals' halos.

The panoply of modern observational techniques includes
slitless dispersed imaging with the VLT+FORS
(\cite[M\'{e}ndez \etal\ 2001]{Mendez01};
\cite[Teodorescu \etal\ 2005]{Teodorescu05}), whose
535 velocities in NGC~4697 remains
by far the largest data-set in an ordinary elliptical.
Three ellipticals have been studied with the PN.S as
a prelude to a large, systematic sample
(\cite[Romanowsky \etal\ 2003]{Romanowsky03}).
PN kinematics have been measured in the
giant Virgo ellipticals M87 and NGC~4472 (M49)
using a ``masked-CDI'' hybrid technique---useful
given pre-imaging data with uncertain astrometry.
Fabry-P\'{e}rot narrow-band scanning has
been used successfully on several galaxies 
(\cite[Sluis \& Williams 2006]{Sluis06})---an
approach which should become more potent with larger telescopes
({\it e.g.}, \cite[Bershady \etal\ 2004]{Bershady04}).

\subsection{Dispersion profiles of elliptical galaxies}

The standard mass indicator in ellipticals is the
projected velocity dispersion profile $\sigma_{\rm p}(R)$,
although it is very non-trivial to translate this into $v_{\rm c}(r)$.
In discussing observed PN dispersions in galaxies, I will also 
mention the dispersions of globular clusters (GCs) when available,
as these are complementary tracers of halo mass. 

In NGC 5128, $\sigma_{\rm p}(R)$ is fairly constant at $\sim 120$~km~s$^{-1}$
for both the PNe and GCs, out to $\sim$~10~$R_{\rm eff}$
(effective radii).
This implies a dark halo, although perhaps not as massive as expected
(\cite[Peng \etal\ 2004]{Peng04}).
In the brightest cluster galaxy NGC~4472, 
the dispersion is also flat for the PNe and GCs to $\sim$~10~$R_{\rm eff}$
(G. Bergond \etal\ in prep).
In M87, the elevated PN and GCs dispersion profiles
reveal the effects of the massive Virgo Cluster core
(Romanowsky \& Kochanek 2001),
but their very different slopes imply in this case a difference
in the orbital properties between the stars and the GCs.
In NGC~1399, the PNe and GCs also show a rising $\sigma_{\rm p}(R)$
(\cite[Arnaboldi \etal\ 1994]{Arnaboldi94};
Y.~Schuberth \etal\ in prep).
This may be evidence of the Fornax Cluster core
(\cite[Saglia \etal\ 2000]{Saglia00}),
or of transient heating from a galaxy interaction
(\cite[Napolitano \etal\ 2002]{Napolitano02}).

\begin{figure}
\centering
\resizebox{6.5cm}{!}{\includegraphics{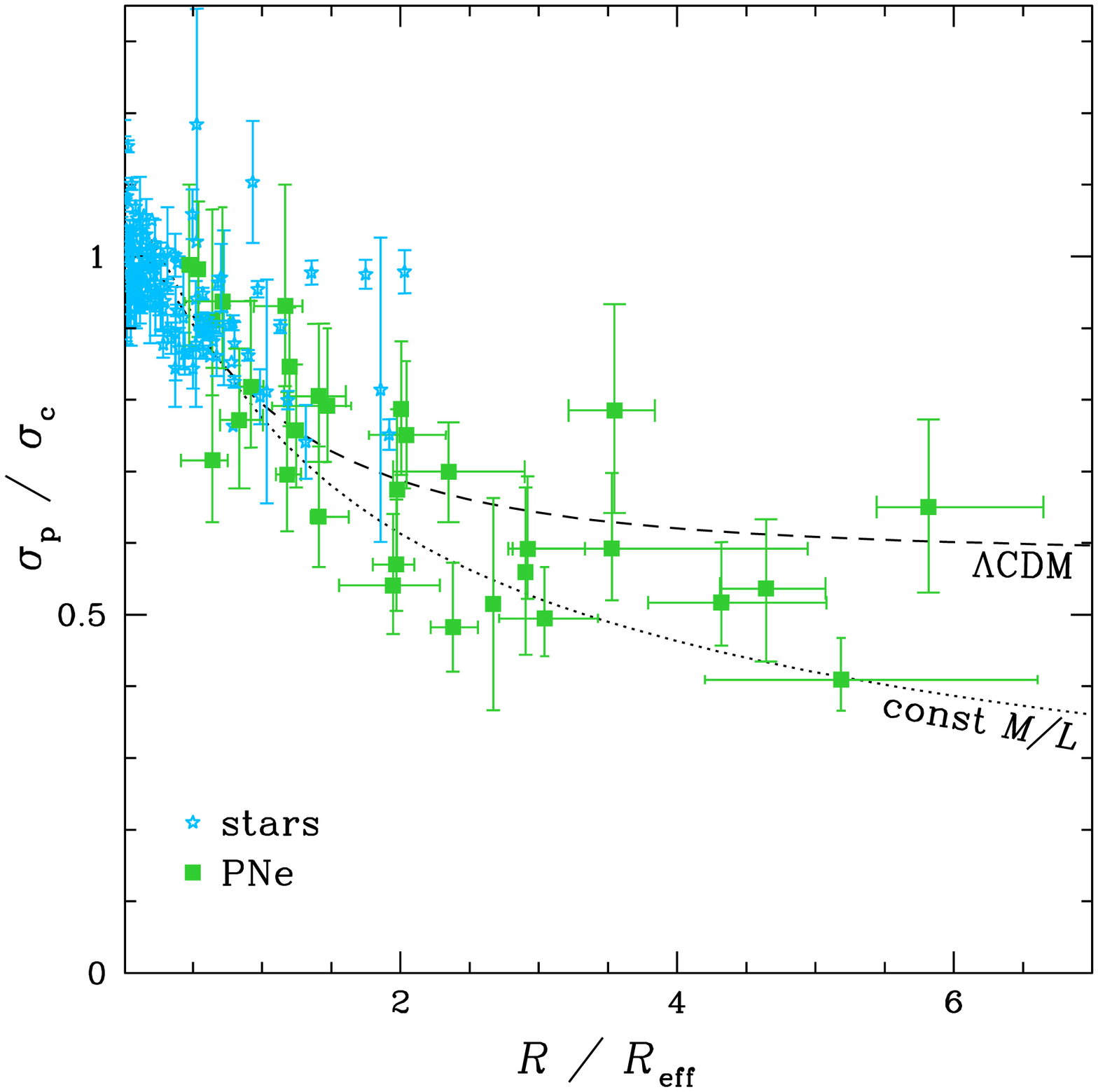} }
\resizebox{6.5cm}{!}{\includegraphics{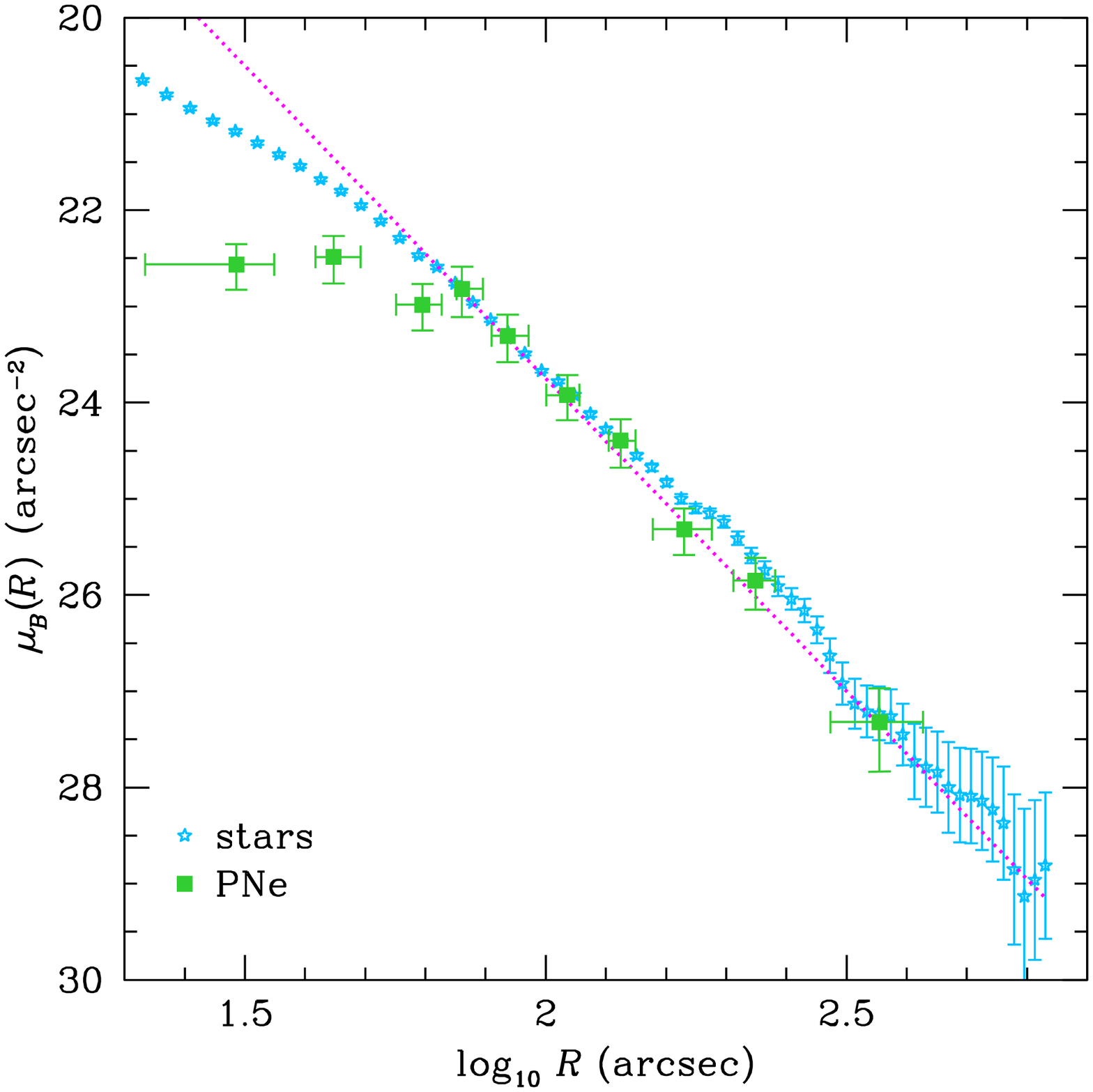} }
\caption[]{
{\it Left:} Rescaled and stacked profiles of projected velocity dispersion
for 5 ordinary elliptical galaxies.
Stellar and PN profiles are shown by different symbols, and
simple model predictions are shown by curves.
{\it Right:}
NGC~3379 surface density profiles.
Points with error bars show the $B$-band stellar photometry
(\cite[Capaccioli \etal\ 1990]{Capaccioli90}),
and the PN number density (with arbitrary normalization).
Inside $75''$, the PN data are incomplete.
The line shows a steeper density profile
that might correspond to young stars formed in a merger 
(see \cite[Dekel \etal\ 2005]{Dekel05}).
}\label{fig:fig2}
\end{figure}

More remarkable behavior is seen in 
the five observed ordinary ellipticals,
where $\sigma_{\rm p}$ declines markedly with radius,
with some galaxy-to-galaxy scatter in the slope
(Fig.~\ref{fig:fig2}, left).
This suggests low DM content, as
first inferred by \cite{Ciardullo93} from PN kinematics in NGC~3379.
NGC~3379 is the only one of these galaxies so far with GC kinematics data
(49 velocities; \cite[Puzia \etal\ 2004]{Puzia04};
\cite[Pierce \etal\ 2006]{Pierce06};
\cite[Bergond \etal\ 2006]{Bergond06}).
The GC halo dispersion declines much more gradually than the PNe, but the implication of this is
not clear (see \S\ref{sec:halom}).

\subsection{Systematic issues in mass estimation}\label{sec:iss}

Before moving on to direct mass results,
I consider a number of complications in deriving elliptical
mass profiles from PN kinematics.
One is that foreground stars and background emission line sources
can masquerade as PNe due to low signal-to-noise or to
limited wavelength coverage.
With current PN.S data, background emitters are rejected based
on resolving the emission in space or wavelength, or
on starkly outlying velocities.

The next issue is the well-known mass-anisotropy degeneracy, whereby
the types of orbits have an effect on the projected dispersions similar
to the impact of the mass distribution.
For example, low $\sigma_p$ can be caused by low mass {\it or} by
radially-biased orbits whose motions are largely in the plane of the sky.
The model curves shown in Fig.~\ref{fig:fig2} (left) are for isotropic
galaxies, but if the PN orbits in the halo have radial anisotropy,
then the $\Lambda$CDM prediction could be substantially lowered. 

Fortunately, this degeneracy can be broken by examining the {\it shape}
of the line-of-sight velocity distribution in addition to its {\it width} (dispersion).
With integrated-light stellar data, this shape is characterized by high-order
Gauss-Hermite polynomials, and for discrete PN velocity data, the shape can
be inferred statistically.
Such analyses can require $\sim$~1000 discrete data points
even under the simplifying assumption of sphericity
(\cite[Merritt \& Saha 1993]{Merritt93})---but
this discouragingly large number is really more relevant for
sparse stellar and galaxy clusters.
With elliptical galaxies, there are additional constraints that
drastically reduce the uncertainties:
the PNe are assumed to be drawn from a well-known spatial distribution,
and the stellar kinematics firmly establish the mass and anisotropy 
characteristics for the central half of the galaxy.

The diffuse stellar component is thus of great importance in exploiting
the PN kinematics, which raises the important question of how well
do the PNe trace the diffuse stellar light?
One effect may arise from radial metallicity gradients, which are thought
to exist in the stellar bodies of ellipticals based on their color gradients.
A metallicity gradient would imply an oxygen gradient, and
thus if the halo PNe had systematically weaker \oiii{} emission,
there would be a bias in the PN detections resulting in lower observed
$\sigma_{\rm p}$.
However, the evidence so far in nearby ellipticals is that
halo metallicities have weakly negative or even {\it positive} gradients
(\cite[Gregg \etal\ 2004]{Gregg04};
\cite[M\'{e}ndez \etal\ 2005]{Mendez05};
\cite[Rejkuba \etal\ 2005]{Rejkuba05};
\cite[Walsh \etal\ 2006]{Walsh06}).
Another potential problem is if the PNe are ``contaminated'' by a young
({\it e.g.} $< 3$ Gyr) population.
However, CMD and PN observations find no evidence for young halo populations
(\cite[Gregg \etal\ 2004]{Gregg04};
\cite[Peng \etal\ 2006]{Peng06}),
and UV fluxes indicate that even in the few early-type galaxies
with recent star formation ($< 1.5$ Gyr),
this contributes only 1--2\% of the stellar mass
(\cite[Yi \etal\ 2005]{Yi05}).

A different approach is to directly test the link between the stars
and PNe:  do they have the same spatial and kinematic distributions in
regions of overlap?
As mentioned earlier, various studies of PNe in elliptical galaxies
have not turned up any indications of population biases.
Even in M31, with its spatially-varying star-formation history,
there are no evident population effects such as correlations between
kinematics and PN luminosity, or variations in the
PN luminosity function with radius.
In NGC~3379, the surface density and kinematics of the stars and the PNe
match up well, although it is difficult to rule out
small differences (see Fig.~\ref{fig:fig2}, right).
The one clear example of a subpopulation 
signature is in NGC~4697,
where an East-West asymmetry of the bright PNe implies an unmixed
stellar component---perhaps corresponding to a recent interaction with another galaxy
(\cite[Sambhus \etal\ 2006]{Sambhus06}).
This feature coincides with the asymmetry of the observational set-up,
so confirmation is needed of the result in this galaxy and in others.

Dynamical models usually assume equilibrium, which will not hold in the
outermost halo regions where crossing times are long.
However, equilibrium should be a good approximation out 
to $\sim$~10~$R_{\rm eff}$ (see \cite[Mamon \etal\ 2006]{Mamon06}).
The final major source of uncertainty is the intrinsic shape of the galaxies.
Modeling flattened systems as spherical can skew mass inferences,
and in particular, a large face-on disk could produce low
velocity dispersions in projection.
However, this particular scenario does not work for NGC~821 and NGC~4697,
which are {\it near-edge-on} ``low DM'' galaxies.

\subsection{Halo masses of elliptical galaxies}\label{sec:halom}

Having considered all these {\it caveats}, we turn to detailed modeling of
NGC~3379, with extensive stellar kinematics 
(\cite[Statler \& Smecker-Hane 1999]{Statler99})
and an initial data set of $\sim$~100 PN velocities
(\cite[Romanowsky \etal\ 2003]{Romanowsky03}).
An ``orbit modeling'' method is used which constrains the
anisotropy directly from the data, assuming spherical symmetry.
The resulting models do indeed have radial anisotropy in the halo,
allowing for more DM than suggested by Fig.~\ref{fig:fig2} (left).
The $B$-band mass-to-light ratio inside 5 $R_{\rm eff}$ is $7.1 \pm 0.6$
(in Solar units),
which is very similar to estimates for the stellar $\Upsilon_*$.
This result is not what one would expect from $\Lambda$CDM theory,
but given the considerable uncertainties in the mass constraints and
in theoretical predictions, $\Lambda$CDM cannot be ruled out.
The other $L^*$ galaxies have not yet been modeled in such detail,
but Jeans model comparisons suggest similar results,
with the exception of NGC~1344.

While signs of DM from PN kinematics may be skimpy,
virial-radius constraints
(from weak gravitational lensing, satellite dynamics, and
galaxy cross-correlations) do demonstrate the presence of 
massive halos around ordinary ellipticals.
Thus the burning question is not whether ellipticals are
``naked'' galaxies without DM, but 
{\it what is the radial distribution of DM} that can
satisfy all these constraints?
This would provide clues about the
physics of galaxy formation and 
about the nature of the DM itself.
A meta-analysis of the available kinematic constraints
suggests that DM halos in ellipticals have low concentrations
(\cite[Napolitano \etal\ 2005]{Napolitano05}),
as found in many other studies of late-type
galaxies and galaxy clusters.
However, these low concentrations may be less of a problem
now as the consensus on the cosmological parameter $\sigma_8$ is 
shifting toward much lower values.
Perhaps this points toward the use of PNe as
precision cosmology tools.

\cite{Dekel05} argue that far from being a problem
for $\Lambda$CDM, the PN data are a natural outcome.
Simulating the formation of elliptical galaxies
via spiral mergers, they find familiarly declining
$\sigma_{p}(R)$ profiles.
The physics of the simulations may not adequately
represent the formation of galaxies in $\Lambda$CDM,
but in any case, the study highlights three major effects
that could contribute to the declines.
The first is varying radial anisotropy, which indeed is included
in our direct modeling of NGC~3379.
The second is that the triaxial structure of the galaxies
causes variations in $\sigma_{\rm p}$ with the viewing angles.
The third is that {\it if} the observed PNe stem from a young
stellar population formed in the merger, then they are not a fair
tracer of the bulk of the stars (see \S\ref{sec:iss}), 
and produce a steeper decline in $\sigma_{\rm p}$.

More definitive statements on PN-based mass profiles obviously await
experiments with triaxial models
({\it e.g.}, \cite[De Lorenzi \etal\ 2006]{DeLorenzi06}).
The large PN.S program sample will also be able to average out viewing-angle effects.
Of further use are independent mass constraints,
such as GCs and X-ray emission.
As mentioned, in NGC~3379, $\sigma_{\rm p}(R)$ is
very different for the PNe and GCs, 
which can be largely explained through different spatial
and anisotropy characteristics.
The GCs do suggest a more massive halo,
but this appears to be contradicted by
the kinematics of a large-radius \hi{} gas ring
implying a remarkably low-mass halo
(\cite[Schneider 1985]{Schneider85}).

\section{Summary}\label{sec:summ}

Planetary nebula are powerful tools for exploring galaxy mass profiles,
and with the advent of new instrumentation, the sample of galaxies
with PN kinematics is now exploding.
In spiral galaxies,
fairly {\it constant} PN velocity dispersions imply
surprisingly large disk mass scale-lengths.
In ordinary ellipticals,
{\it declining} PN dispersions suggest low-concentration dark matter halos.
An important issue still needing clarification for dynamical models is
the connection between PNe and their parent stellar population.

\begin{acknowledgments}
I would like to thank Kim Herrmann \& Robin Ciardullo for providing their plots,
the Scientific Organizing Committee for their invitation, and
the IAU for travel support.
\end{acknowledgments}

\end{document}